\documentclass[12pt]{article}

\PassOptionsToPackage{hidelinks,unicode,hypertexnames=false}{hyperref}

\usepackage[margin=1in]{geometry}
\usepackage[utf8]{inputenc}
\usepackage[T1]{fontenc}
\usepackage{booktabs}
\usepackage{graphicx}
\usepackage{amsmath}
\usepackage{tabularx}
\usepackage{array}
\usepackage{longtable}
\usepackage{float}
\usepackage{nameref}
\usepackage{hyperref}
\usepackage{silence}
\newcolumntype{Y}{>{\raggedright\arraybackslash}X}

\hypersetup{
  colorlinks=true,
  linkcolor=black,
  citecolor=black,
  urlcolor=black
}

\newcommand{\seeref}[1]{\hyperref[#1]{\nameref{#1}}}

\title{Thermal fluctuations and electronic readout\\using a moving-coil movement}

\author{
Angelo Maria Sabatini\\
The BioRobotics Institute\\
Scuola Superiore Sant'Anna\\
Pisa, Italy\\
\texttt{angelo.sabatini@santannapisa.it}
}
\date{}

\begin{document}

\maketitle

\begin{abstract}
The moving-coil meter provides a simple context in which to model deterministic electromechanical dynamics, thermal fluctuations, and electronic readout. The movement is a linear electromechanical system in which electrical resistance and intrinsic mechanical damping act as two independent dissipative channels, each accompanied by thermal fluctuations. The stored energy provides a Lyapunov function for stability, and the algebraic Lyapunov equation yields the stationary covariance in closed form. The result exhibits equipartition among the energy terms; damping is shown to reshape the temporal memory of the fluctuations without changing their steady-state variance, giving a thermal floor of a few picoamperes for a representative movement. Exact stochastic integration reproduces the stationary covariance and autocorrelation structure; the covariance is also recovered by spectral integration. Embedding the movement in an active current-drive readout shifts attention to the noise of the complete readout, which a design-stage estimate expresses as a few parts per million of full scale, growing as the input sensitivity increases. The example illustrates that measurement sensitivity and resolution cannot be considered independently. The treatment is aimed at master’s-level students in physics with some background in statistical physics and stochastic processes, and basic familiarity with analog electronics.

\end{abstract}

\section{Introduction}

A moving-coil meter converts an electrical quantity into the angular deflection of a coil suspended in a permanent magnetic field. Its operating principle is elementary, but asking what ultimately limits the smallest detectable deflection leads beyond the deterministic response of the instrument. Electrical resistance and mechanical damping dissipate energy, and at thermal equilibrium the same dissipative channels generate fluctuations. The moving-coil meter therefore provides a simple physical system in which electromechanical transduction, dissipation, thermal noise, and measurement performance can be examined within a common model.

The development of the instrument closely parallels that of the physical ideas needed to describe these fluctuations. At the heart of the meter, the moving-coil movement, historically associated with the d’Arsonval galvanometer, was patented by d’Arsonval and Deprez in 1881 and refined a few years later by Weston \cite{weston_galvanometer}, becoming a reference for sensitive electrical measurements. Langevin’s treatment of Brownian motion in 1908 introduced a stochastic description of thermally driven motion \cite{lemons1997}; years later, Johnson experimentally observed thermal fluctuations associated with electrical resistance, while Nyquist provided their statistical-mechanical interpretation \cite{johnson1928,nyquist1928}.

Electrical resistance in the movement’s circuit had already been recognized as a design parameter for electromagnetic damping \cite{wenner_1916_galvanometers}, while thermal fluctuations of the moving coil were known before the Nyquist–Johnson formulation \cite{ornstein1927,niss2016}. The subsequent Ornstein–Uhlenbeck formulation of Brownian motion \cite{uhlenbeck_ornstein1930} provides a framework that, still today, allows the moving-coil movement to be treated as a linear dynamical system driven by thermal fluctuations.

The moving-coil movement is well suited here to bringing together several ideas that students often encounter separately in measurement science, transducer modelling, and statistical physics. The electrical and mechanical parts of it exchange energy through bilateral electromechanical coupling, while resistance and intrinsic mechanical damping provide distinct dissipative channels. Fluctuation--dissipation associates a thermal driving term with each of them \cite{pecseli2000}.

In a state-space modelling approach, the stored energy is a natural candidate for a Lyapunov function in the stability analysis of the underlying deterministic dynamics. The stationary covariance of the corresponding Ornstein--Uhlenbeck process can then be obtained by solving the associated algebraic Lyapunov equation (ALE). The resulting covariance makes equipartition explicit \cite{pecseli2000}, and it will turn out that dissipation shapes the temporal correlation structure of the fluctuations without changing their equilibrium variance. According to the theory of stationary processes, the same results can alternatively be obtained through the Wiener--Khinchin theorem \cite{brown_hwang} by integrating the power and cross-spectral densities to recover the complete stationary covariance. The accompanying simulation code also allows the exact stochastic integration, stationary covariance, and autocorrelation structure to be reproduced numerically and compared with an independent frequency-domain calculation, offering a self-contained computational exercise in stochastic state-space modelling.

The overall treatment is intended for master's students in physics with some previous exposure to statistical physics and stochastic processes. Only basic familiarity with circuits based on operational amplifiers (OP-AMPs) is assumed for the analysis of the active current drive coupled to the moving-coil movement. OP-AMP circuits are commonly encountered in undergraduate physics laboratories \cite{papanikolaou2015} and measurement courses as basic conditioning and actuation stages within a measurement chain \cite{dunn2005}.

The entire modeling flow will be surveyed, from the deterministic electromechanical model and the intrinsic thermal floor of the movement to the design-stage noise analysis of the complete OP-AMP-based readout. The noise will be referred to the input (RTI) and expressed relative to input full scale \cite{franco_design}. 

The moral we take from the paper is that measurement sensitivity and resolution cannot be considered independently: increasing the input sensitivity of the readout also increases the noise relative to full scale.

\section{Anatomy of the moving-coil movement}

By its defining response law, the moving-coil movement is current-sensitive. A current-carrying coil rotates against a restoring torque in the uniform radial field produced by a permanent magnet. Its motion may be opposed by electromagnetic damping, when the back electromotive force (EMF) is allowed to drive current in the electrical circuit, as well as by intrinsic damping (eddy-current or oil-bath). With a steady current through the coil, the pointer attached to it settles where the magnetic torque balances the mechanical restoring torque of the suspension.

\begin{figure}[!ht]
\centering
\includegraphics[width=\columnwidth]{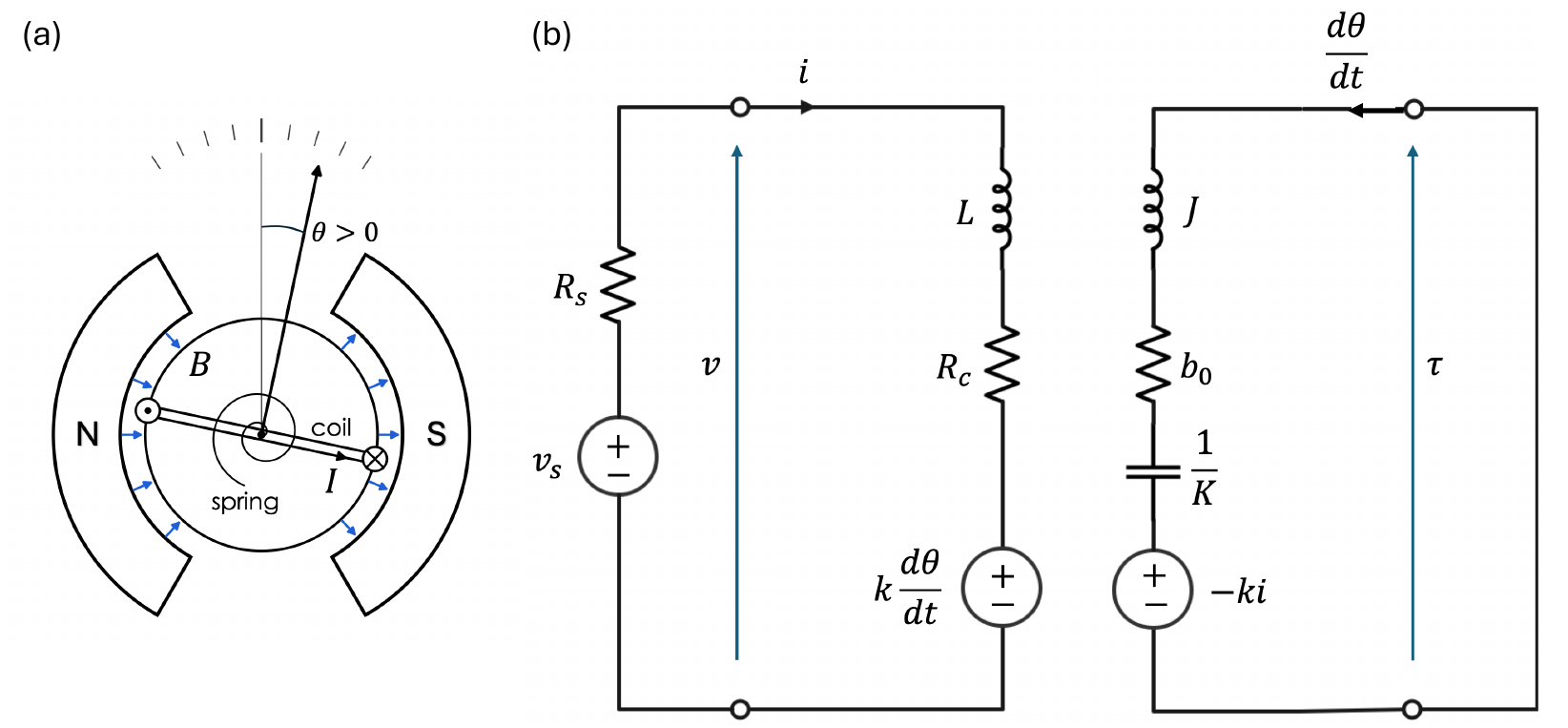}
\caption{(a) Schematic cross section of the moving-coil movement. (b) Two-port representation of its coupled electrical and mechanical dynamics in the impedance analogy. The mechanical port is shown short-circuited, corresponding to $\tau=0$.}
\label{fig-two-port}
\end{figure}

Figure~\ref{fig-two-port} depicts the resulting two-port structure: the moving-coil assembly couples an electrical port, at which current $i$ flows in response to a voltage $v$, to a mechanical port, at which the coil turns through $\theta$ in response to electromechanical coupling and possibly to an external torque $\tau$. In the time domain, the coupled dynamics take the form
\begin{equation}
\begin{cases}
v = R_c i + L\dfrac{di}{dt} + k\dfrac{d\theta}{dt},\\[6pt]
J\dfrac{d^2\theta}{dt^2} + b_0\dfrac{d\theta}{dt} + K\theta = ki+\tau.
\end{cases}
\label{eq-coupled-time}
\end{equation}
Here $R_c$ and $L$ are the coil's resistance and inductance, $J$ is the moment of inertia of the moving assembly, $b_0$ its intrinsic damping, and $K$ the torsional stiffness of the suspension; $\tau$ is any additional torque applied directly to the shaft, retained here for generality. The same constant $k$ appears in both equations, coupling the electrical port's flow variable $i$ into the mechanical port's effort equation and the mechanical port's flow variable $d\theta/dt$ into the electrical port's effort equation: this reciprocal, power-conserving cross-coupling between conjugate effort-flow pairs is precisely what characterizes a gyrator, the two-port element proper to electromechanical transduction, as distinct from the effort-to-effort, flow-to-flow coupling of a transformer. At the mechanical port, however, the shaft of the movement carries only its own suspension and pointer in ordinary operation: no deterministic external torque is applied to it, so $\tau=0$. In generalized network language, the mechanical port is therefore short-circuited: its effort variable vanishes while its flow variable $d\theta/dt$ remains free to evolve.

Collecting $\theta, d\theta/dt$ and $i$ into the state vector $\mathbf{x}$, the short-circuited system of Eq.~(\ref{eq-coupled-time}) takes the explicit state-space form
\begin{equation}
\dfrac{d\mathbf{x}}{dt} =
\underbrace{\begin{bmatrix}
0 & 1 & 0\\
-K/J & -b_0/J & k/J\\
0 & -k/L & -R_c/L
\end{bmatrix}}_A \mathbf{x} +
\underbrace{\begin{bmatrix}
0\\ 0\\ 1/L
\end{bmatrix}}_{\mathbf{b}}\,v.
\label{eq-state-space-det}
\end{equation}
The state $\mathbf{x}$ evolves under the drift matrix $A$, driven by the electrical input $v$ through $\mathbf{b}$.

Equation~(\ref{eq-state-space-det}) can be Laplace-transformed from zero initial conditions. With $\mathrm{X}(s)=(\mathrm{\Theta},\mathrm{\Omega},\mathrm{I})^{\mathsf T}$ the transform of the state vector and $\mathrm{V}(s)$ that of the electrical input, we write
\begin{equation}
\mathrm{X}(s)=(s\mathbf{I}-A)^{-1}\,\mathbf{b}\,\mathrm{V}(s),
\label{eq-state-space-laplace}
\end{equation}
with $\mathbf{I}$ the $3\times3$ identity matrix. Equation~(\ref{eq-state-space-laplace}) casts the movement's dynamics in the form standard in linear systems and circuit theory: it gives at once the three transfer functions relating $\theta$, $d\theta/dt$, and $i$ to the driving voltage $v$, with $\det(s\mathbf{I}-A)$, a cubic polynomial in $s$, determining their common denominator. For example:
\begin{equation}
\dfrac{\Theta(s)}{V(s)} = \dfrac{k}{JLs^3 +(R_cJ +Lb_0)s^2 + (R_cb_0 + k^2 + LK)s + R_cK}.
\label{eq-HthetaV}
\end{equation}
The mechanical driving-point impedance of the moving assembly is
\begin{equation}
Z_{mm}(s) = Js + b_0 + \dfrac{K}{s}.
\label{eq-Zmm}
\end{equation}
Through the gyrator, this impedance is reflected to the electrical port as $k^2/Z_{mm}(s)$. The driving-point impedance seen at the coil terminals is therefore
\begin{equation}
Z_{ee}(s) = R_c + sL + \dfrac{k^2}{Z_{mm}(s)} = R_c + sL + \dfrac{k^2}{Js + b_0 + K/s}.
\label{eq-Zee}
\end{equation}
Thus the mechanical dynamics appear directly in the electrical driving-point impedance: the coil cannot, in general, be represented by its resistance and inductance alone. Equation~(\ref{eq-Zee}) is the impedance-domain counterpart of the bilateral coupling already expressed by Eq.~(\ref{eq-coupled-time}).

The exact electromechanical dynamics are therefore third order. In the usual operating regime of the moving-coil movement, however, the electrical time scale of the bare movement is much shorter than the mechanical one,
\begin{equation}
\dfrac{L}{R_c}\ll\sqrt{\dfrac{J}{K}}.
\label{eq-scale-separation}
\end{equation}
On the mechanical time scale, the inductive term can then be neglected. When $v$ is constant to set the DC operating point $I_{\mathrm{DC}}$, the current can be written $i=I_{\mathrm{DC}}+\delta i$, with the back EMF producing the incremental current
\begin{equation}
\delta i\simeq-\dfrac{k}{R_c}\dfrac{d\theta}{dt},
\label{eq-current-quasistatic}
\end{equation}
since the coil resistance is the only resistance at the electrical port. The mechanical dynamics then reduce to
\begin{equation}
J\dfrac{d^2\theta}{dt^2} + \left(b_0 + \dfrac{k^2}{R_c}\right)\dfrac{d\theta}{dt} +  K\theta = kI_{\mathrm{DC}}.
\label{eq-reduced-dynamics}
\end{equation}
The resulting electromagnetic damping coefficient is
\begin{equation}
b_{\mathrm{em}}=\dfrac{k^2}{R_c}.
\label{eq-bem}
\end{equation}
Writing the magnetic torque as $\kappa(\theta)\,i$, with $\kappa(\theta)$ the coupling coefficient between the electrical and mechanical ports of the movement, and the restoring torque as $K\theta$, equilibrium requires $\kappa(\theta_{\mathrm{DC}})\,I_{\mathrm{DC}}=K\theta_{\mathrm{DC}}$. For the ideal geometry considered here, $\kappa$ is a constant $k$ throughout the operating angular range by construction --- the radial field seen by the coil does not depend on $\theta$ \cite{bentley_measurement_systems}. Therefore the static sensitivity of the movement at the operating point $I_{\mathrm{DC}}$ is
\begin{equation}
S(I_{\mathrm{DC}}) = \left.\dfrac{d\theta}{di}\right|_{i=I_{\mathrm{DC}}} = \dfrac{k}{K},
\label{eq-static-sensitivity}
\end{equation}
and is independent of $I_{\mathrm{DC}}$ throughout the operating range.

The electrical termination does not alter this static sensitivity, but it can strongly affect the transient response. When the movement is connected to an external circuit, the resistance through which the back EMF can drive current is no longer necessarily the coil resistance $R_c$ alone. Let $R$ denote the total effective resistance \emph{seen by the coil}. Under the same time-scale separation used above, the electromagnetic damping coefficient becomes
\begin{equation}
b_{\mathrm{em}}=\dfrac{k^2}{R},
\label{eq-bem-general}
\end{equation}
and $R$ replaces $R_c$ also in Eq.~\ref{eq-reduced-dynamics}.

Introducing the natural angular frequency $\omega_n=\sqrt{K/J}$ and the damping ratio $\zeta$ in the usual second-order form, the reduced dynamics give
\begin{equation}
\zeta = \dfrac{b_0+k^2/R}{2\sqrt{JK}}.
\label{eq-damping-ratio}
\end{equation}
So far, we have considered a scenario where the movement is voltage driven. If the voltage source is ideal ($R_s=0$ and $R=R_s+R_c=R_c$), the electromagnetic damping takes its maximum value, $\zeta_{\mathrm max} = (b_0+k^2/R_c)/(2\sqrt{JK})$. If $\zeta_{\mathrm max} > 1$, a properly chosen external resistance $R_s$, connected in series with the movement, can reduce the damping to the critical condition $\zeta=1$. The corresponding critical-damping external resistance (CDRX) is
\begin{equation}
R_{\mathrm{CDRX}} = \dfrac{k^2}{2\sqrt{JK}-b_0}-R_c.
\label{eq-CDRX}
\end{equation}
On the other hand, the movement can also be current driven. If the current source is ideal, the electromagnetic damping vanishes, since the back EMF is not allowed to change the current flowing in the coil. In this case, only the intrinsic damping can be active, with $\zeta_{\mathrm min}=b_0/(2\sqrt{JK})$. Critical damping can nevertheless be restored by connecting a resistance $R_p$ in parallel across the terminals of the movement. When the ideal current source is suppressed, the back EMF drives current through the loop formed by $R_c$ and $R_p$, so that the effective resistance is $R=R_c+R_p$. By Norton--Thévenin equivalence, the critical-damping condition therefore requires $R_p=R_{\mathrm{CDRX}}$.

Thus voltage and current drive can realize the same finite damping condition through equivalent resistive terminations, whereas their ideal limits lead to opposite electromagnetic-damping conditions. What determines the damping is the resistance of the closed path available to the back EMF.

The same total resistance $R$ that sets the damping ratio also affects the frequency range over which the movement can follow a changing input. For the reduced second-order model, the $-3$-dB bandwidth is
\begin{equation}
B = f_n\sqrt{(1 - 2\zeta^2)+ \sqrt{(1 - 2\zeta^2)^2 + 1}},
\label{eq-bandwidth}
\end{equation}
which, for fixed $f_n=\omega_n/(2\pi)$, is a monotonically decreasing function of $\zeta$. For critical damping, $\zeta=1$, Eq.~(\ref{eq-bandwidth}) reduces to
\begin{equation}
B = f_n\sqrt{\sqrt{2}-1} \approx 0.644\,f_n.
\end{equation}
Table~\ref{tab-canonical-parameters} fixes a canonical set of parameters for a representative movement, used for every numerical example in the remainder of this paper. $J$, $L$, and $k$ follow from the geometry of a single coil design; $K$ is chosen to give the specified full-scale deflection, here $I_{\mathrm{FS}}=\pm100\,\mu\text{A}$ and $\theta_{\mathrm{FS}}=\pm60^\circ$, through Eq.~(\ref{eq-static-sensitivity}). The bare-coil resistance is $R_c=15.28\,\Omega$, while $b_0$ represents the intrinsic mechanical damping. Throughout the numerical examples we choose critical damping, $\zeta=1$, which fixes the required electrical termination through Eq.~(\ref{eq-CDRX}).

\begin{table}
\caption{Canonical parameters and derived quantities used for the numerical example.}
\label{tab-canonical-parameters}
\renewcommand{\arraystretch}{0.88}
\begin{tabular}{lll}
\hline\hline
Quantity & Value & Origin \\
\hline
$R_c$ & $15.276\,\Omega$ & coil geometry \\
$L$ & $7.271\times10^{-4}\,\text{H}$ & coil geometry \\
$k$ & $6.283\times10^{-3}\,\text{N m/A}$ & coil geometry \\
$J$ & $3.149\times10^{-8}\,\text{kg m}^2$ & coil geometry \\
$b_0$ & $4.170\times10^{-8}\,\text{N m s}$ & intrinsic damping \\
$K$ & $6.000\times10^{-7}\,\text{N m/rad}$ & from full-scale specification \\
\hline
$R_{\mathrm{CDRX}}$ & $154.004\,\Omega$ & Eq.~(\ref{eq-CDRX}) \\
$R$ & $169.280\,\Omega$ & critical damping, $\zeta=1$ \\
$S$ & $1.047\times10^4\,\text{rad/A}$ & Eq.~(\ref{eq-static-sensitivity}) \\
$f_n$ & $0.695\,\text{Hz}$ & $\sqrt{K/J}/(2\pi)$ \\
$B$ & $0.447\,\text{Hz}$ & Eq.~(\ref{eq-bandwidth}), $\zeta=1$ \\
\hline\hline
\end{tabular}
\end{table}

It is worth noting that the damping budget adopted for the numerical example follows a classical rule of thumb. Wenner reports a representative critically damped movement in which about 15\% of the total damping is intrinsic and the remaining 85\% is supplied electromagnetically by the external circuit \cite{wenner_1916_galvanometers}. For the numerical example of Table~\ref{tab-canonical-parameters}, $b_0/(2\sqrt{JK})=0.152$, so that intrinsic and electromagnetic damping account for 15.2\% and 84.8\% of the critical value, respectively.

\section{Thermal fluctuations and damping}

Because the system considered here is linear, its deterministic response and its response to thermally induced stochastic fluctuations separate by superposition. The thermal fluctuations can therefore be studied with the deterministic excitation set to zero, so that the equilibrium position is $\theta_{\mathrm{DC}}=0$. Section II also identified $R$, the total effective resistance closing the loop around the coil. At this operating point the movement is not mechanically or electrically silent: $R$ and the intrinsic damping channel both dissipate energy and, by the same token, both drive it. Sitting at the same ambient temperature $T$, each channel drives its own fluctuating force: the resistor a voltage $e(t)$ in series with it, with the white-noise correlation established by Johnson and Nyquist \cite{johnson1928,nyquist1928} $\langle e(t)e(t')\rangle=2k_BTR\,\delta(t-t')$, where $k_B$ is Boltzmann's constant; the mechanical channel a fluctuating torque $\tau_{\mathrm{th}}(t)$, statistically independent of $e(t)$, with the fluctuation-dissipation form familiar from Langevin's equation, \cite{lemons1997} $\langle\tau_{\mathrm{th}}(t)\tau_{\mathrm{th}}(t')\rangle=2k_BTb_0\,\delta(t-t')$.

In state-space form, the stochastic dynamics become
\begin{equation}
\dfrac{d\mathbf{x}}{dt} =
\underbrace{\begin{bmatrix}
0 & 1 & 0\\
-K/J & -b_0/J & k/J\\
0 & -k/L & -R/L
\end{bmatrix}}_A \mathbf{x} + 
\underbrace{\begin{bmatrix}
0\\ 1/J\\ 0
\end{bmatrix}}_{\mathbf{g}_1}\,\tau_{\mathrm{th}} + 
\underbrace{\begin{bmatrix}
0\\ 0\\ 1/L
\end{bmatrix}}_{\mathbf{g}_2}\,e.
\label{eq-state-space-stoch}
\end{equation}
This is precisely what is now called an Ornstein--Uhlenbeck process \cite{uhlenbeck_ornstein1930}: a linear system driven by Gaussian white noise and confined by a stable drift matrix $A$ to a stationary Gaussian distribution, with two independent fluctuating forces associated with the two dissipative channels. The diffusion matrix $D$ built from the two noise intensities fixed above is
\begin{equation}
D = \mathbf g_1\mathbf g_1^{\mathsf T}\cdot 2k_BTb_0 + \mathbf g_2\mathbf g_2^{\mathsf T}\cdot 2k_BTR = 
\begin{bmatrix} 0 & 0 & 0\\ 0 & 2k_BTb_0/J^2 & 0\\ 0 & 0 & 2k_BTR/L^2
\end{bmatrix}.
\label{eq-diffusion-matrix}
\end{equation}
The matrix $D$ can also be expressed in terms of the matrix
\begin{equation}
\Gamma = 
\begin{bmatrix}
0 & 0 & 0\\
0 & b_0/J^2 & 0\\
0 & 0 & R/L^2
\end{bmatrix}
\end{equation}
as $D = 2k_BT\,\Gamma$.

Writing $\omega=d\theta/dt$ for the angular velocity, the energy stored in the coupled system is
\begin{equation}
V(\mathbf{x}) = \dfrac{1}{2} \mathbf{x}^\mathsf T
\underbrace{\begin{bmatrix}
K & 0 & 0\\ 0 & J & 0\\ 0 & 0 & L
\end{bmatrix}}_{M} \mathbf{x} = \dfrac{1}{2}\left(K\theta^2 + J\omega^2 + Li^2\right).
\label{eq-energy}
\end{equation}
It takes little algebra to recognize that 
\begin{equation}
AM^{-1}=S-\Gamma,
\label{eq-S-Gamma}
\end{equation}
where
\begin{equation}
S = \begin{bmatrix}0 & 1/J & 0\\ -1/J & 0 & k/(JL)\\ 0 & -k/(JL) & 0\end{bmatrix}
\end{equation}
$S$ is antisymmetric: it moves energy between the mechanical and electrical degrees of freedom without creating or destroying it, the same coupling responsible for the torque and the back EMF above. $\Gamma$ is symmetric and positive semidefinite, and it collects every dissipative channel present --- here, the two associated with $b_0$ and $R$.

An important question is whether the deterministic flow $d{\mathbf x}/dt = A\mathbf x$, obtained by switching both sources of thermal noise off, actually settles rather than growing without bound. The energy $V(\mathbf x)$ answers this question directly. Because $M$ is diagonal, $M^{-1}A^{\mathsf T}=(AM^{-1})^{\mathsf T}=-S-\Gamma$, so along the deterministic flow
\begin{equation}
\dfrac{dV}{dt} = \mathbf x^{\mathsf T}M\,\dfrac{d\mathbf x}{dt} = \mathbf x^{\mathsf T}MA\,\mathbf x = \dfrac12\mathbf x^{\mathsf T}(MA + A^{\mathsf T}M) \mathbf x = -\mathbf x^{\mathsf T}M\Gamma M\, \mathbf x = -b_0\omega^2 - Ri^2\le0,
\label{eq-lyapunov-decay}
\end{equation}
using $MA+A^{\mathsf T}M=-2M\Gamma M$, itself a direct consequence of Eq.~(\ref{eq-S-Gamma}) with $S$ antisymmetric and $\Gamma$ symmetric. Energy can only decrease, and only through the two channels that dissipate it.

For $b_0>0$ and $R>0$, equality $dV/dt = 0$ requires both $\omega = 0$ and $i = 0$; on that set the mechanical equation then gives $\theta = 0$. The same conclusion also holds when the intrinsic damping is absent, $b_0 = 0$, provided that $R > 0$ and $k \neq 0$. In that case $dV/dt = 0$ requires only $i = 0$, but a trajectory can remain in this set only if $di/dt = -k\omega/L = 0$, which implies $\omega = 0$; the mechanical equation then gives $\theta = 0$. Thus the origin is the largest invariant set contained in $\{dV/dt=0\}$ in either case. By LaSalle's invariance principle, the origin is therefore globally asymptotically stable, and $A$ is Hurwitz. The deterministic system thus converges to a unique stable equilibrium. For the stochastic system, this is exactly the condition classical theory requires to guarantee a unique stationary Gaussian distribution in the presence of the corresponding thermal noise sources \cite{arnold1974,khasminskii2012}.

With existence secured, the stationary covariance $\Sigma = \langle \mathbf{x} \mathbf{x}^{\mathsf T} \rangle$ is governed by the ALE
\begin{equation}
A\Sigma + \Sigma A^{\mathsf T} + D = 0.
\label{eq-ale}
\end{equation}
Because $M^{-1}A^{\mathsf T} = -S-\Gamma$, the ansatz $\Sigma = k_BTM^{-1}$ gives
\begin{equation}
A\Sigma+\Sigma A^{\mathsf T} = k_BT\left[(S-\Gamma)+(-S-\Gamma)\right] = -2k_BT\Gamma = -D,
\label{eq-ansatz-check}
\end{equation}
satisfying Eq.~(\ref{eq-ale}) exactly. The stationary covariance is therefore diagonal,
\begin{equation}
\Sigma = k_BT\,M^{-1} = \mathrm{diag}\!\left(\dfrac{k_BT}{K}, \dfrac{k_BT}{J}, \dfrac{k_BT}{L}\right).
\label{eq-equipartition}
\end{equation}
Thus the three quadratic degrees of freedom satisfy equipartition,
\begin{equation}
\dfrac{1}{2}K\langle\theta^2\rangle = \dfrac{1}{2}J\langle\omega^2\rangle = \dfrac{1}{2}L\langle i^2\rangle = \dfrac{1}{2}k_BT.
\label{eq-equipartition-energy}
\end{equation}
Moreover, since the stationary state is jointly Gaussian and its covariance matrix is diagonal, the three state variables are statistically independent. Neither the damping coefficient $b_0$ nor the resistance $R$, nor how the total dissipation is split between the two dissipative channels, appears in $\Sigma$. In the frequency domain, changing $b_0\geq0$ or $R$ redistributes the thermal fluctuations over frequency, but leaves their integrated variance unchanged. The Appendix provides an independent frequency-domain cross-check, recovering the complete stationary covariance element by element through spectral integration à la Wiener--Khinchin \cite{brown_hwang}.

With the parameters of the numerical example indicated in Table~\ref{tab-canonical-parameters}, the RMS angular jitter of the pointer at rest and its equivalent input RMS thermal-noise current, respectively, follow directly from Eqs.~(\ref{eq-static-sensitivity}, \ref{eq-equipartition}). At ambient temperature $T = 300\,\mathrm{K}$:
\begin{equation}
\theta_{\mathrm {RMS}}=\sqrt{\dfrac{k_BT}{K}} = 8.31\times10^{-8}\,\mathrm{rad}, \qquad
I_{\mathrm{th, RMS}} =\dfrac{K}{k}\theta_{\mathrm{RMS}} = \dfrac{\sqrt{k_BTK}}{k} = 7.93\,\mathrm{pA}.
\label{eq-theta-rms-ith}
\end{equation}
The latter is about seven orders of magnitude below the $\pm100\,\mu\mathrm{A}$ full-scale current.

\subsection{The memory of thermal fluctuations}
\label{sec-thermal-memory}

The equipartition result above fixes the steady-state variance of each state, but not how long a thermal fluctuation retains memory of itself. For the reduced second-order model of Eq.~(\ref{eq-damping-ratio}), the stationary autocorrelation of the deflection angle is
\begin{equation}
C_\theta(\tau)=\Sigma_{\theta\theta}\,e^{-\zeta\omega_n|\tau|}\left[\cos(\omega_d\tau)+\dfrac{\zeta\omega_n}{\omega_d}\sin(\omega_d|\tau|)\right],
\label{eq-acf-theory}
\end{equation}
for $\zeta<1$, with $\omega_d=\omega_n\sqrt{1-\zeta^2}$. At critical damping, $\zeta=1$, the double-root limit is
\begin{equation}
C_\theta(\tau)=\dfrac{k_BT}{K}\,e^{-\omega_n|\tau|}\left(1+\omega_n|\tau|\right),
\label{eq-acf-critical}
\end{equation}
where $C_\theta(0)=\Sigma_{\theta\theta}=k_BT/K$, as required by Eq.~(\ref{eq-equipartition}).

\begin{figure}[!ht]
\centering
\includegraphics[width=0.8\columnwidth]{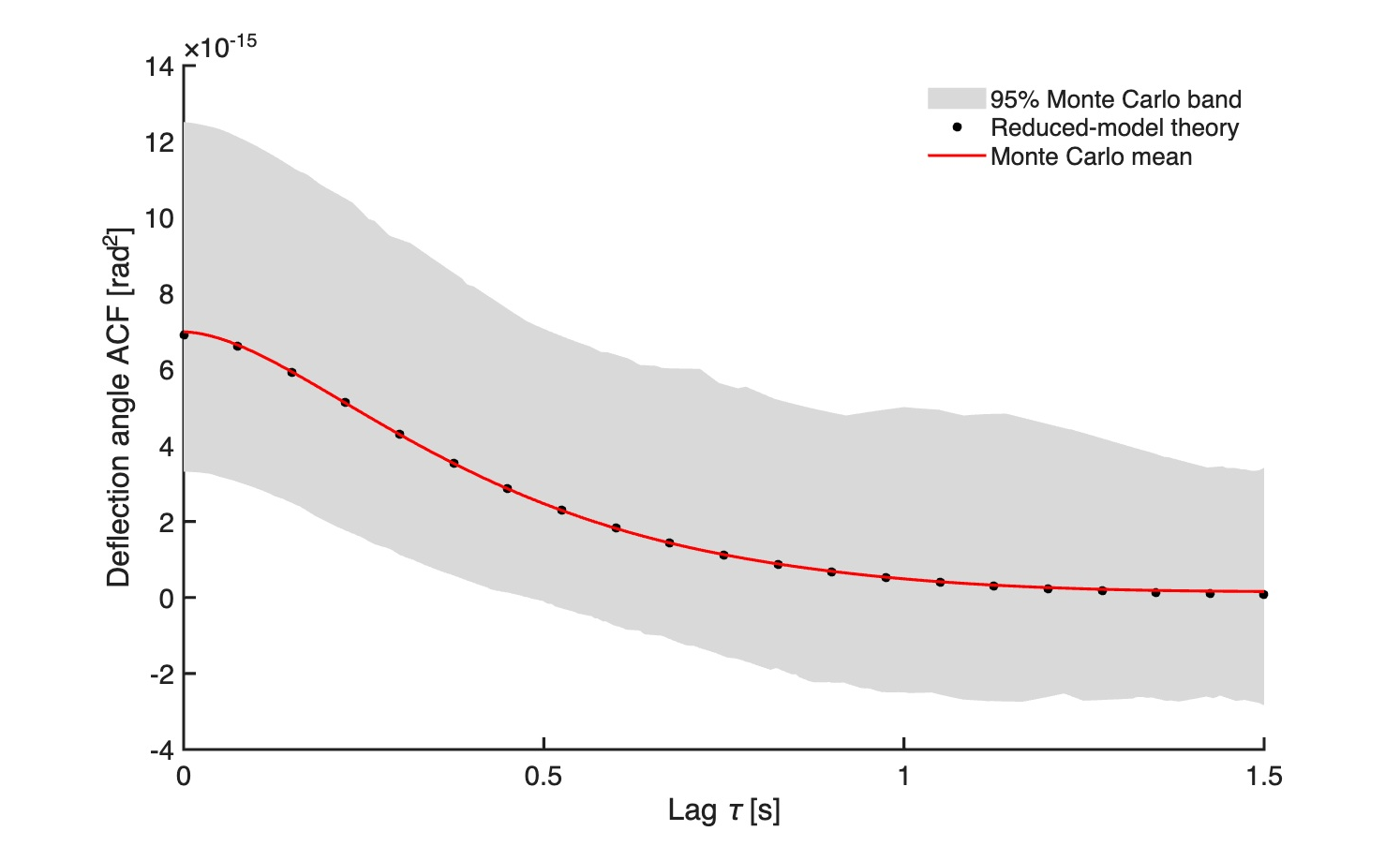}
\caption{Stationary autocorrelation of the deflection angle at the critically damped design point of Table~\ref{tab-canonical-parameters}. Black points show the reduced-model prediction of Eq.~(\ref{eq-acf-critical}); the colored curve and shaded region show the mean and 95\% band, respectively, over 500 independent Monte Carlo realizations of the full third-order stochastic system. At zero lag, $C_\theta(0)=k_BT/K$, in agreement with Eq.~(\ref{eq-equipartition}).}
\label{fig-acf-verification}
\end{figure}

The reduction from the full third-order stochastic system to Eq.~(\ref{eq-acf-critical}) rests on the same separation between the electrical and mechanical time scales used in Sec.~II. For the parameters of Table~\ref{tab-canonical-parameters}, $L/R=4.30\times10^{-6}\,\mathrm{s}$ whereas $1/\omega_n=\sqrt{J/K}=2.29\times10^{-1}\,\mathrm{s}$, a separation of more than four orders of magnitude. Figure~\ref{fig-acf-verification} tests the reduced result against 500 independent Monte Carlo realizations of the full third-order stochastic system, including both thermal-noise sources. Each realization is initialized directly from the equilibrium Gaussian distribution with covariance $\Sigma=k_BT M^{-1}$ and propagated exactly between sampled states using the Van Loan discretization \cite{brown_hwang}, without a burn-in interval or an Euler--Maruyama approximation \cite{arnold1974}. The Monte Carlo mean closely follows the reduced-model prediction over the full range of lags shown. Simulation code and parameters are provided in the accompanying software repository \cite{sabatini2026software}.

\section{Reading out the deflection: an active current-drive loop}

Section~II introduced an example movement designed to be critically damped. We now turn to a widely used circuit for driving the movement and reading out its deflection. The topology is a non-inverting voltage-to-current converter \cite{franco_design}: the input voltage $V_{\mathrm{in}}$ drives the non-inverting terminal, the sense resistor $R_{\mathrm{sense}}$ connects the inverting terminal to ground, and a resistor $R_{\mathrm{add}}$ closes the feedback path through the movement. Under the usual virtual-short approximation, the inverting terminal is held at $V_{\mathrm{in}}$. Since no other path leaves that node, the same current $I_{\mathrm{out}}$ flows through $R_{\mathrm{sense}}$ and the feedback path,
\begin{equation}
I_{\mathrm{out}}=\dfrac{V_{\mathrm{in}}}{R_{\mathrm{sense}}}.
\label{eq-current-drive}
\end{equation}

\begin{figure}[!ht]
\centering
\includegraphics[width=0.8\textwidth]{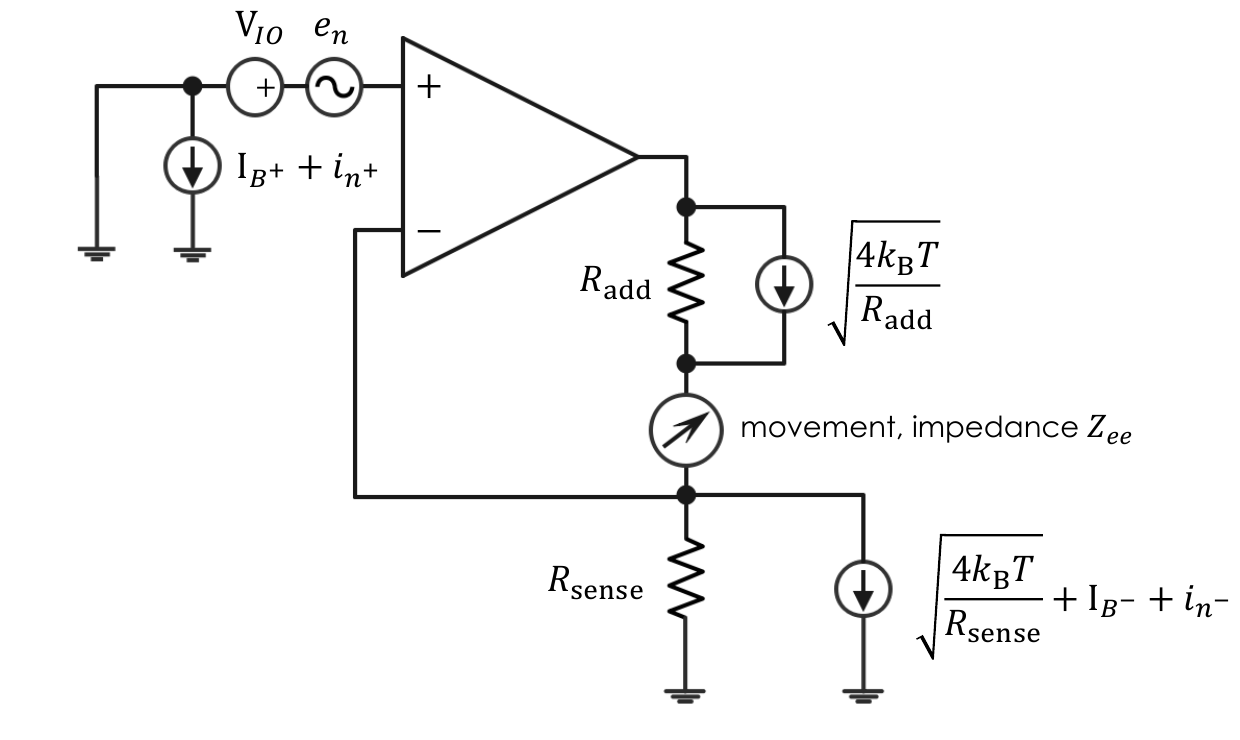}
\caption{Voltage-to-current converter with the error and noise sources considered in the analysis. The movement is terminated by the critical-damping resistance $R_p$.}
\label{fig-current-drive-noise}
\end{figure}

The value chosen for $R_{\mathrm{add}}$ in this exercise is $100\,\mathrm{k\Omega}$. As discussed in Sec.~II, the movement is terminated by the critical-damping resistance $R_p$ connected in parallel across its terminals, as represented by the movement symbol in Fig.~\ref{fig-current-drive-noise}. By Norton--Thévenin equivalence, the current drive and $R_p$ correspond at the movement terminals to a voltage source $R_p I_{\mathrm{out}}$ in series with $R_p$, so that the back EMF sees the total resistance $R_c+R_p$ required for critical damping. The movement presents the impedance $Z_{ee}$ of Eq.~(\ref{eq-Zee}), with $|Z_{ee}(i\omega_n)|\approx1\,\mathrm{k\Omega}$, which is small compared with $R_{\mathrm{add}}$. The feedback impedance may therefore be approximated by $R_{\mathrm{add}}$, both in evaluating the loop gain and the noise gain $NG$,
\begin{equation}
NG\simeq1+\dfrac{R_{\mathrm{add}}}{R_{\mathrm{sense}}}.
\label{eq-noise-gain}
\end{equation}

Figure~\ref{fig-current-drive-noise} shows the same circuit with the error and noise sources included in the analysis.

The numerical example uses an OPA277 precision operational amplifier. Its data sheet gives an input offset voltage $V_{IO} = 10\,\mu\mathrm{V}$, input bias current $I_B = 0.5\,\mathrm{nA}$, white input-voltage-noise density $e_{nw} = 8\,\mathrm{nV}/\sqrt{\mathrm{Hz}}$, and white input-current-noise density $i_{nw} = 0.2\,\mathrm{pA}/\sqrt{\mathrm{Hz}}$; the voltage- and current-noise spectra exhibit approximate $1/f$ corner frequencies $f_{ce} = 3\,\mathrm{Hz}$ and $f_{ci} = 20\,\mathrm{Hz}$, respectively.\cite{opa277_datasheet} Offset and bias currents are retained in Fig.~\ref{fig-current-drive-noise} to complete the circuit model but are not included in the noise budget below. Their nominal values, tolerances, and thermal drifts could readily be propagated through the same circuit, but they belong to an error and drift budget rather than to the stationary noise calculation considered here.

The virtual-short approximation makes the noise analysis particularly transparent. We work in the frequency domain with power spectral densities (PSDs): each noise source is propagated through the corresponding transfer function, so that its contribution to the output PSD is weighted by the squared magnitude of that transfer function. We assume the noise sources to be mutually uncorrelated, so that their PSD contributions can be added directly \cite{franco_design}. The sources fall naturally into four groups.

At the non-inverting input, the random voltage noise $e_n$ appears in exactly the same way as the signal $V_{\mathrm{in}}$. Its contribution to the output current PSD is therefore
\begin{equation}
S_{I,e_n}(f) = \dfrac{S_{e_n}(f)}{R_{\mathrm{sense}}^2}.
\label{eq-en-current}
\end{equation}
The current noise $i_{n^+}$ produces no voltage in the present idealized circuit because, with an ideal voltage source suppressed, the non-inverting input is shorted to ground. As noted above, the DC quantities $V_{IO}$ and $I_{B^+}$ are excluded from the noise analyis.

At the inverting input, the amplifier current noise $i_{n^-}$ and the Johnson current noise of $R_{\mathrm{sense}}$ are injected directly into the feedback path. Under the same virtual-short condition they flow into the feedback path and hence contribute directly to the output current. The latter has the white PSD
\begin{equation}
S_{I,R_{\mathrm{sense}}}(f)=\dfrac{4k_BT}{R_{\mathrm{sense}}}.
\label{eq-Rsense-johnson}
\end{equation}
The Johnson current generated by $R_{\mathrm{add}}$, in contrast, does not change the output current in the ideal closed-loop limit: the amplifier output simply supplies the additional voltage required to preserve the current imposed by Eq.~(\ref{eq-current-drive}). Its contribution is therefore absent from the present output-current noise budget.

Finally, the movement and its damping termination $R_p$ constitute the thermal system already analyzed in Sec.~III. Their combined equilibrium contribution is therefore not added source by source again here. Its integrated input-referred thermal noise is
\begin{equation}
I_{\mathrm{th,RMS}} = \dfrac{\sqrt{k_BTK}}{k},
\label{eq-Ithermal-readout}
\end{equation}
as obtained in Eq.~(\ref{eq-theta-rms-ith}). This contribution is statistically independent of the noise added by the readout electronics and will be compared with the corresponding integrated electronic noise below.

The electronics contribution to the output-current noise is consequently described by the PSD
\begin{equation}
S_{I,\mathrm{el}}(f) = \dfrac{S_{e_n}(f)}{R_{\mathrm{sense}}^2} + S_{i_n}(f) + \dfrac{4k_BT}{R_{\mathrm{sense}}},
\label{eq-electronics-current-psd}
\end{equation}
where
\begin{equation}
S_{e_n}(f) = e_{nw}^2\left(1+\dfrac{f_{ce}}{f}\right), \qquad S_{i_n}(f) = i_{nw}^2\left(1+\dfrac{f_{ci}}{f}\right).
\label{eq-opa-noise-model}
\end{equation}
Three values of $R_{\mathrm{sense}}$ are considered: $10$, $100$, and $1000\,\Omega$. With $R_{\mathrm{add}}=100\,\mathrm{k\Omega}$, their corresponding noise gains are 10{,}001, 1{,}001, and 101. Even for the largest noise gain, the OPA277 gain-bandwidth product of $1\,\mathrm{MHz}$ places the electrical closed-loop bandwidth near $100\,\mathrm{Hz}$, far above the mechanical frequencies relevant here. The amplifier noise may therefore be regarded as applied to the movement without additional filtering by the electrical loop; it is the much slower mechanical response that sets the effective noise bandwidth.

For the critically damped second-order dynamics considered here, the equivalent noise bandwidth is
\begin{equation}
B_{\mathrm{eq}}=\dfrac{\pi}{4}f_n.
\label{eq-ENBW}
\end{equation}
The white part of Eq.~(\ref{eq-electronics-current-psd}) therefore gives immediately
\begin{equation}
I_{\mathrm{el,RMS,w}}^2 =
\left(
\dfrac{e_{nw}^2}{R_{\mathrm{sense}}^2}
+i_{nw}^2
+\dfrac{4k_BT}{R_{\mathrm{sense}}}
\right)B_{\mathrm{eq}}.
\label{eq-white-noise-rms}
\end{equation}
The $1/f$ terms require one additional step. Because their spectra diverge as $f\rightarrow0$, a low-frequency cutoff must be introduced. For a finite observation time $T_{\mathrm{obs}}$, we take $f_L=1/T_{\mathrm{obs}}$. With $T_{\mathrm{obs}}=5\,\mathrm{s}$, $f_L=0.2\,\mathrm{Hz}$, and the noise contribution is obtained from
\begin{equation}
I_{\mathrm{el,RMS}}^2=\int_{f_L}^{\infty}S_{I,\mathrm{el}}(f)|H(f)|^2\,df.
\label{eq-noise-integral}
\end{equation}
For the white-noise terms, the equivalent noise bandwidth gives the result directly; when the $1/f$ terms are retained, Eq.~(\ref{eq-noise-integral}) is evaluated numerically.

The total RMS current noise is finally
\begin{equation}
I_{\mathrm{tot,RMS}} = \sqrt{I_{\mathrm{th,RMS}}^2 + I_{\mathrm{el,RMS}}^2},
\label{eq-total-current-noise}
\end{equation}
where the first term is the equilibrium floor of the movement itself and the second is the additional noise introduced by the readout electronics. This separation is useful physically: the first is fixed by the fluctuation-dissipation result of Sec.~III, whereas the second can be traded against sensitivity through $R_{\mathrm{sense}}$.

The role of $R_{\mathrm{sense}}$ becomes particularly transparent when the noise is referred back to the input voltage of the converter. Since $V_{\mathrm{in}}=I_{\mathrm{out}}R_{\mathrm{sense}}$, the full-scale movement current $I_{\mathrm{FS}}=100\,\mu\mathrm{A}$ corresponds to input ranges of $1$, $10$, and $100\,\mathrm{mV}$ for $R_{\mathrm{sense}}=10$, $100$, and $1000\,\Omega$, respectively. Decreasing $R_{\mathrm{sense}}$ therefore selects a progressively more sensitive voltage scale.

The same transformation applies to the noise budget. An RMS current noise $I_{\mathrm{RMS}}$ corresponds to the input-referred voltage noise
\begin{equation}
V_{\mathrm{RMS}}=R_{\mathrm{sense}}I_{\mathrm{RMS}}.
\label{eq-input-referred-noise}
\end{equation}
The electronic current noise increases strongly as $R_{\mathrm{sense}}$ is reduced, but its input-referred voltage remains nearly constant because the dominant contribution is the amplifier voltage noise $e_n/R_{\mathrm{sense}}$. The thermal floor of the movement behaves differently: $I_{\mathrm{th,RMS}}$ is fixed by the movement itself, so that $V_{\mathrm{th,RMS}}=R_{\mathrm{sense}}I_{\mathrm{th,RMS}}$ decreases in direct proportion to $R_{\mathrm{sense}}$.

The resulting full-scale-to-noise dynamic range may be written
\begin{equation}
\mathrm{DR}_{\mathrm{dB}}=20\log_{10}\left(\frac{V_{\mathrm{FS}}}{V_{\mathrm{tot,RMS}}}\right).
\label{eq-dynamic-range}
\end{equation}
The same ratio may also be expressed directly as noise in parts per million of full scale. Table~\ref{tab-noise-budget} summarizes the resulting trade-off.

\begin{table}[!ht]
\centering
\caption{Input-referred noise and full-scale-to-noise performance for the three measurement scales. The white-noise results use $B_{\mathrm{eq}}=\pi f_n/4$; the full calculation includes the $1/f$ terms with $T_{\mathrm{obs}}=5\,\mathrm{s}$. The dynamic range and noise in parts per million of full scale are calculated from the total noise including the $1/f$ contributions.}
\label{tab-noise-budget}
\begin{tabular}{rrrrrrr}
\hline\hline
$R_{\mathrm{sense}}$ & $V_{\mathrm{FS}}$ & $V_{\mathrm{th,RMS}}$ & \multicolumn{2}{c}{$V_{\mathrm{tot,RMS}}$} & $\mathrm{DR}$ & Noise/FS \\
$(\Omega)$ & & & White & $1/f$ included & (dB) & (ppm) \\
\hline
10   & $1\,\mathrm{mV}$   & $0.079\,\mathrm{nV}$ & $5.92\,\mathrm{nV}$  & $13.45\,\mathrm{nV}$ & 97.4  & 13.45 \\
100  & $10\,\mathrm{mV}$  & $0.793\,\mathrm{nV}$ & $6.04\,\mathrm{nV}$  & $13.49\,\mathrm{nV}$ & 117.4 & 1.349 \\
1000 & $100\,\mathrm{mV}$ & $7.93\,\mathrm{nV}$  & $10.34\,\mathrm{nV}$ & $15.82\,\mathrm{nV}$ & 136.0 & 0.158 \\
\hline\hline
\end{tabular}
\end{table}

The three scales expose the trade-off directly. Their total input-referred noise corresponds to approximately $13.5$, $1.35$, and $0.16$ parts per million of full scale: nearly one decade of improvement for each decade increase in $R_{\mathrm{sense}}$. At the most sensitive scale, the $100\,\mu\mathrm{A}$ full-scale current corresponds to only $1\,\mathrm{mV}$ at the converter input. Here the movement itself contributes only $0.079\,\mathrm{nV}$ RMS when referred to the input, whereas the complete readout gives $13.45\,\mathrm{nV}$ RMS when the $1/f$ noise is included; the observable noise floor is therefore set overwhelmingly by the readout electronics. As the input range increases, the electronic contribution becomes progressively less important relative to full scale, until the intrinsic thermal floor of the movement begins to emerge. These values are design-stage estimates, obtained from the numerical model and the data-sheet specifications of a representative precision operational amplifier, rather than measurements on a real readout. 

\section{Conclusion}

The moving-coil meter provides a simple context in which deterministic electromechanical dynamics, thermal fluctuations, and electronic readout can be followed within a single modelling framework. The stored energy provides a natural Lyapunov function for establishing stability, while the same energy matrix enters the ALE, whose closed-form solution gives the stationary covariance. The resulting covariance exhibits equipartition among the elastic, kinetic, and magnetic energy terms and gives an equivalent-current thermal floor of about $8\,\mathrm{pA}$ for the example considered here, independent of the electrical and mechanical damping coefficients. Changing the damping reshapes the temporal memory of the fluctuations, but not their integrated steady-state variance.

Once the movement is embedded in an active current-drive circuit --- a voltage--current converter, here --- the relevant quantity is no longer the thermal noise of the movement alone, but the noise of the complete readout. Referring this noise back to the voltage--current converter input makes the trade-off particularly transparent: changing the sense resistance by decades changes the input full-scale voltage by the same factor, while the electronic and movement noise contributions scale differently when referred to the input. For the three scales considered here, the total input-referred noise corresponds to about $13.5, 1.35$, and $0.16$ parts per million of full scale, respectively, from the most to the least sensitive scale. The intrinsic thermal floor of the movement therefore begins to emerge only at the least sensitive scale. These values are design-stage estimates, obtained from the numerical model and the data-sheet specifications of a representative precision operational amplifier, rather than measurements on a real readout. They help quantify one aspect of the noise budget: offset, bias currents, and thermal or long-term drift were not accounted for in the present analysis.

The numerical treatment provides complementary checks of the analytical results. Exact stochastic integration reproduces the stationary covariance and the predicted autocorrelation structure, while the frequency-domain calculation in the Appendix independently recovers the covariance matrix through spectral integration. The agreement among state-space analysis, time-domain simulation, and spectral calculation provides a direct consistency check of the stochastic model. The subsequent noise-budget calculation follows the standard input-referred noise analysis used for OP-AMP-based circuits, extending the same quantitative description from the intrinsic fluctuations of the movement to the complete readout.

\section*{A spectral cross-check}
\label{app-spectral}

Section III obtained the stationary covariance directly in state space from the ALE. Here we recover the same result independently in the frequency domain, by integrating the power and cross-spectral densities generated by the two thermal-noise sources.

With zero initial conditions, the Laplace-domain counterpart of Eq.~(\ref{eq-state-space-stoch}) is
\begin{equation}
\mathrm{X}(s) = (sI-A)^{-1}\left[\mathbf g_1\,\mathrm{T}_{\mathrm{th}}(s) + \mathbf g_2\,\mathrm{E}(s)\right].
\label{eq-laplace-stoch}
\end{equation}
All six transfer functions from the two noise sources to the three state variables share the physical characteristic polynomial
\begin{equation}
\Delta(s) = JLs^3 + (RJ + Lb_0)s^2 + (Rb_0 + k^2 + LK)s + RK =JL\det(sI-A).
\label{eq-charpoly}
\end{equation}
Equation~(\ref{eq-laplace-stoch}) can then be written as
\begin{equation}
\mathrm{X}(s) = \mathbf H_\tau(s)\,\mathrm{T}_{\mathrm{th}}(s) + \mathbf H_e(s)\,\mathrm{E}(s),
\end{equation}
where
\begin{equation}
\mathbf H_\tau(s) = \dfrac{1}{\Delta(s)}
\begin{bmatrix}
Ls + R\\ s(Ls + R)\\ -ks
\end{bmatrix},
\qquad
\mathbf H_e(s) = \dfrac{1}{\Delta(s)}
\begin{bmatrix}
k\\ ks\\ Js^2 + b_0s + K
\end{bmatrix}.
\label{eq-noise-transfer-vectors}
\end{equation}
The two thermal sources are statistically independent and have the two-sided power spectral densities $2k_BT b_0$ and $2k_BT R$. The spectral-density matrix of the state is therefore
\begin{equation}
S_x(\Omega) = 2k_BT b_0\,\mathbf H_\tau(i\Omega)\mathbf H_\tau^\dagger(i\Omega) + 2k_BT R\,\mathbf H_e(i\Omega)\mathbf H_e^\dagger(i\Omega),
\label{eq-state-psd}
\end{equation}
where $^\dagger$ denotes conjugate transpose and the Wiener--Khinchin relation gives
\begin{equation}
\Sigma = \dfrac{1}{2\pi}\int_{-\infty}^{\infty}S_x(\Omega)\,d\Omega.
\label{eq-wiener-khinchin}
\end{equation}
Writing $Q(\Omega) = |\Delta(i\Omega)|^2$, the twelve contributions to the six independent elements of $\Sigma$ are given below, where $\operatorname{Re}$ denotes the real part of a complex quantity; only the real part of each cross-spectral density contributes to the corresponding zero-lag covariance.
\begin{align}
S_{\theta\theta}^{(\tau)}& = \dfrac{2k_BT b_0(R^2 + L^2\Omega^2)}{Q},&
S_{\theta\theta}^{(e)}& = \dfrac{2k_BT Rk^2}{Q},\nonumber\\
S_{\omega\omega}^{(\tau)}& = \dfrac{2k_BT b_0\Omega^2(R^2 + L^2\Omega^2)}{Q},&
S_{\omega\omega}^{(e)}& = \dfrac{2k_BT Rk^2\Omega^2}{Q},\nonumber\\
S_{ii}^{(\tau)}& = \dfrac{2k_BT b_0k^2\Omega^2}{Q},&
S_{ii}^{(e)}& = \dfrac{2k_BT R[(K - J\Omega^2)^2 + b_0^2\Omega^2]}{Q},\nonumber\\
\operatorname{Re} S_{\theta\omega}^{(\tau)}& = 0,&
\operatorname{Re} S_{\theta\omega}^{(e)}& = 0,\nonumber\\
\operatorname{Re} S_{\theta i}^{(\tau)}& = -\dfrac{2k_BT b_0kL\Omega^2}{Q},&
\operatorname{Re} S_{\theta i}^{(e)}& = \dfrac{2k_BT Rk(K-J\Omega^2)}{Q},\nonumber\\
\operatorname{Re} S_{\omega i}^{(\tau)}& = -\dfrac{2k_BT b_0kR\Omega^2}{Q},&
\operatorname{Re} S_{\omega i}^{(e)}& = \dfrac{2k_BT Rkb_0\Omega^2}{Q}.
\label{eq-spectral-contributions}
\end{align}
The diagonal terms in Eq.~(\ref{eq-spectral-contributions}) can be evaluated analytically using the Cauchy--Phillips formulas for rational integrals \cite{brown_hwang}. For
\begin{equation}
I_n=\dfrac{1}{\pi}\int_0^\infty
\dfrac{c(i\Omega)c(-i\Omega)}
{d(i\Omega)d(-i\Omega)}\,d\Omega,
\end{equation}
where
\begin{equation}
c(s) = c_{n-1}s^{n-1} + \cdots + c_1s + c_0, \qquad d(s) = d_ns^n + \cdots + d_1s + d_0,
\end{equation}
the cubic case is
\begin{equation}
I_3 =\dfrac{c_2^2d_0d_1 + (c_1^2 - 2c_0c_2)d_0d_3 + c_0^2d_2d_3}{2d_0d_3(d_1d_2 - d_0d_3)}.
\label{eq-phillips-I3}
\end{equation}
For the present system, $d(s)=\Delta(s)$, so that
\begin{equation}
d_3 = JL,\qquad d_2 = RJ + Lb_0,\qquad d_1 = Rb_0 + k^2 + LK,\qquad d_0 = RK.
\label{eq-phillips-coefficients}
\end{equation}
The coefficients $d_i$ are therefore common to all six diagonal spectral contributions in Eq.~(\ref{eq-spectral-contributions}), whereas the numerator coefficients $c_i$ depend on the state variable and on the thermal-noise source. Factoring out the corresponding noise intensity, $2k_BT b_0$ or $2k_BT R$, the six cases are
\begin{equation}
\begin{array}{c|ccc}
\text{Contribution} & c_0 & c_1 & c_2\\
\hline
S_{\theta\theta}^{(\tau)} & R & L & 0\\
S_{\theta\theta}^{(e)}  & k & 0 & 0\\
S_{\omega\omega}^{(\tau)} & 0 & R & L\\
S_{\omega\omega}^{(e)} & 0 & k & 0\\
S_{ii}^{(\tau)} & 0 & -k & 0\\
S_{ii}^{(e)} & K & b_0 & J
\end{array}
\label{eq-phillips-numerators}
\end{equation}
Substitution of the six sets of numerator coefficients into Eq.~(\ref{eq-phillips-I3}) gives, on defining
\begin{equation}
\mathcal D = JR^2b_0 + JRk^2 + KL^2b_0 + LRb_0^2 + Lb_0k^2,
\end{equation}
\begin{align}
I_3^{(\theta,\tau)} &= \dfrac{JR^2 + KL^2 + LRb_0}{2K\mathcal D},&
I_3^{(\theta,e)} &= \dfrac{k^2(JR + Lb_0)}{2KR\mathcal D},\nonumber\\
I_3^{(\omega,\tau)} &= \dfrac{JR^2 + KL^2 + LRb_0 + Lk^2}{2J\mathcal D},&
I_3^{(\omega,e)} &= \dfrac{k^2}{2\mathcal D},\nonumber\\
I_3^{(i,\tau)} &= \dfrac{k^2}{2\mathcal D},&
I_3^{(i,e)} & = \dfrac{JR^2b_0 + JRk^2 + KL^2b_0 + LRb_0^2}{2LR\mathcal D}.
\label{eq-phillips-results}
\end{align}
Restoring the two thermal-noise intensities, the diagonal elements of the stationary covariance are therefore
\begin{equation}
\Sigma_{xx} = 2k_BT b_0 I_3^{(x,\tau)} + 2k_BT R I_3^{(x,e)}, \qquad x\in\{\theta,\omega,i\},
\end{equation}
which reduce exactly to
\begin{equation}
\Sigma_{\theta\theta} = \dfrac{k_BT}{K},\qquad
\Sigma_{\omega\omega} = \dfrac{k_BT}{J},\qquad
\Sigma_{ii} = \dfrac{k_BT}{L}.
\label{eq-spectral-variances}
\end{equation}
The cross-covariances exhibit a different structure. For $\theta$ and $\omega$, the real parts of both cross-spectral contributions vanish separately,
\begin{equation}
\operatorname{Re}S_{\theta\omega}^{(\tau)}(\Omega) = \operatorname{Re}S_{\theta\omega}^{(e)}(\Omega)=0,
\end{equation}
because, for either noise source, the transfer function to $\omega$ is $i\Omega$ times that to $\theta$. The corresponding cross-spectra are therefore purely imaginary, and
\begin{equation}
\Sigma_{\theta\omega} = 0.
\end{equation}

For $\omega$ and $i$, neither contribution vanishes separately. Instead,
\begin{equation}
\operatorname{Re}S_{\omega i}^{(\tau)}(\Omega) = -\dfrac{2k_BT b_0kR\Omega^2}{Q(\Omega)}, \qquad
\operatorname{Re}S_{\omega i}^{(e)}(\Omega) = \dfrac{2k_BT Rkb_0\Omega^2}{Q(\Omega)},
\end{equation}
so that the two contributions cancel pointwise at every frequency. Hence
\begin{equation}
\Sigma_{\omega i} = 0.
\end{equation}
The remaining cross-covariance, $\Sigma_{\theta i}$, vanishes only after integration. Summing the two corresponding contributions in Eq.~(\ref{eq-spectral-contributions}) gives
\begin{equation}
\operatorname{Re}S_{\theta i}(\Omega) = \dfrac{2k_BT k\left[RK-(RJ+b_0L)\Omega^2\right]}{Q(\Omega)}.
\end{equation}
Since the integrand is even, the Wiener--Khinchin integral can again be expressed in terms of Cauchy--Phillips integrals. The two terms correspond respectively to the numerator coefficients
\begin{equation}
(c_0, c_1, c_2) = (1, 0, 0), \qquad (c_0, c_1, c_2) = (0, 1, 0),
\end{equation}
for which Eq.~(\ref{eq-phillips-I3}) gives
\begin{equation}
I_3^{(0)} = \dfrac{d_2}{2d_0(d_1d_2-d_0d_3)}, \qquad
I_3^{(2)} = \dfrac{1}{2(d_1d_2-d_0d_3)}.
\end{equation}
It follows that
\begin{align}
\Sigma_{\theta i} &= 2k_BT k\left[RK\, I_3^{(0)} - (RJ + b_0L)I_3^{(2)}\right]\nonumber\\ &= \dfrac{k_BT k}{d_1d_2-d_0d_3}
\left(\dfrac{RK\,d_2}{d_0} - RJ-b_0L\right) = 0,
\label{eq-spectral-theta-i}
\end{align}
because $d_0 = RK$ and $d_2 = RJ + Lb_0$.

Collecting the three variances and the three cross-covariances, the spectral calculation therefore gives
\begin{equation}
\Sigma =
k_BT
\begin{bmatrix}
1/K & 0 & 0\\
0 & 1/J & 0\\
0 & 0 & 1/L
\end{bmatrix} = k_BT M^{-1}.
\end{equation}
Thus the frequency-domain calculation recovers, element by element, the same stationary covariance obtained directly from the ALE in Sec.~III. In particular, it reproduces both the three equipartition variances and the vanishing of all three cross-covariances.

\end{document}